\documentclass[twocolumn,showpacs,preprintnumbers,amsmath,amssymb]{revtex4}

\usepackage{graphicx}
\usepackage{dcolumn}
\usepackage{bm}

\newcommand{\mean}[1]{\langle #1 \rangle}

\begin{document}


\title{Methods of calculation of a friction coefficient: \\ Application to
the nanotubes}

\author{J. Servantie}
\author{P. Gaspard}
\affiliation{
 Center for Nonlinear Phenomena and Complex Systems\\
Universit\'e Libre de Bruxelles, Code Postal 231, Campus Plaine, 1050
Brussels, Belgium.}

\date{\today}

\begin{abstract}
In this work we develop theoretical and numerical methods of calculation of
a dynamic friction coefficient. The theoretical
method  is based on an adiabatic approximation which allows us to express
the dynamic friction coefficient
in terms of the time integral of the autocorrelation function of the force
between both sliding objects.
The motion of the objects and the autocorrelation function can be numerically
calculated by molecular-dynamics simulations.
We have successfully applied these methods to the
evaluation of the dynamic friction coefficient of the relative motion of
two concentric carbon nanotubes.
The dynamic friction coefficient is shown to increase with the temperature.
\end{abstract}


\maketitle

Thanks to recent developments in nanotechnology, the hope is high to build
mechanical devices
on the scale of the nanometer.  For this purpose, it is important to
determine the mechanical
properties and, especially, the friction forces in such nanodevices.
In this work, we are going to focus
on systems of carbon nanotubes. Our particular interest is for a system
that Cumings and Zettl \cite{Zettl} observed experimentally
with a TEM. They fixed an edge of a multiwalled nanotube to a surface and
opened the other edge, they then extracted inner layers
from the core for several nanometers and released them. They observed a
full retraction of the inner layers and furthermore,
they could conclude that the multiwalled nanotubes are self cleaning since
amorphous carbon due to the opening of the edge are
inside the tube and also do not have any wear, structural change or fatigue
after several extraction and retraction processes.
This experiment suggest us that the multiwalled nanotubes can be promising
systems for future nanometric mechanical parts
such as springs, gears or even motors.

A short time after the work of Cumings and Zettl, Zheng and Jiang
\cite{Zheng} estimated the frequency of the
oscillations in this system to be of the order of GHz. This result is also
very interesting since in the macroscopic world
moving parts with such frequencies does not exist at the present time. But
as in the macroscopic world moving parts have
friction forces which hinder the motion and dissipate energy. 
We hence have to
know the importance of these forces before
conceiving such devices. Our work will thus focus on, firstly methods of
calculation of the friction and secondly an
application of these theories to the multiwalled nanotubes.

The plan of the paper will be as follows, we are first going to introduce
the theoretical framework for the
description of the mechanics of nanotubes. We will then solve the
classical equations of motion and calculate the dynamic friction coefficient
by molecular-dynamics simulation and by the autocorrelation-function method
developed
by Jarzynski, Berry and Robbins \cite{Jarzynski,Berry}.

\begin{figure}[h]
\begin{center}
\includegraphics[scale=0.3]{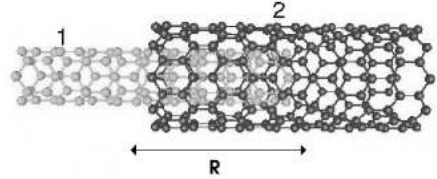}
\caption{\label{sliding} Sliding nanotubes.}
\end{center}
\end{figure}

We have depicted in Fig. \ref{sliding} two sliding nanotubes. $R$ is the
distance between the centers of mass of
each nanotube.
The Hamiltonian of the system of two nanotubes is given by
\begin{equation}
H=T_1+T_2+V_{\rm TB}^{(1)}+V_{\rm TB}^{(2)}+ \frac{1}{2}\sum_{ij}
V_{\rm LJ}\left(\mathbf{r}_i^{(1)}-\mathbf{r}_j^{(2)}\right)
\end{equation}
where $T_1$ and $T_2$ are respectively the total kinetic energies of the
inner and outer tubes, $V_{\rm TB}^{(1)}$
and $V_{\rm TB}^{(2)}$ are the total Tersoff-Brenner potentials of both
tubes. We use the set of parameters which optimizes
the  geometric structure \cite{Brenner}. The last term is the interaction
potential between the tubes, which is taken as a
6-12  Lennard-Jones potential calculated by Lu and Yang \cite{Lu} and widely 
used for molecular-dynamics simulations of 
nanotubes \cite{Zheng,Lu,Lu2,Legoas}.

The calculation of the friction coefficient will be done by using an
adiabatic approximation since there are two
different time scales in the system. The fast time scale is the one of the
vibrations of the atoms
which is of the order of the $fs$ and the slow time scale is the one of the
motion of the mass centers of the tubes which
is of the  order of the $ps$ in our case. Thanks to this difference in time
scales,
we can use a method developed by Jarzynski, Berry and Robbins
\cite{Jarzynski,Berry} which is based on the fact that the following
quantity is an adiabatic invariant
\cite{Ott,Goldstein},
\begin{equation}
\Omega(E,\epsilon t)=\int d\mathbf{z}\ \ \theta
\left[E-H(\mathbf{z},\epsilon t)\right]
\label{invariant}
\end{equation}
where $\epsilon$ is a small parameter, E the energy of the fast system and 
$H$ is an \textit{ergodic adiabatic Hamiltonian}. This means that
$H$ slowly evolves with time and that, at any instantaneous time, $H$
produces trajectories which explore their energy shells ergodically and
chaotically.
$\mathbf{z}=(\mathbf{q},\mathbf{p})$ represents the fast phase-space
variables. The smallness of the parameter $\epsilon$ guarantees that 
the motion of vibration is fast compared to the motion of the mass centers. 
The other important assumption is that the
force on the slow system can be expressed as an average over the fast
degrees of freedom since the
time scale of the fast system is much smaller. We hence write the force as
\begin{equation}
\mathbf{F}=-\int d\mathbf{z} \; \rho \; \mathbf{\nabla} H
\label{forcemean}
\end{equation}
where $\rho$ is the probability density of the fast system and
$\mathbf{\nabla}=\partial/\partial \mathbf{R}$ is the
derivative with respect to the slow variable. In our
specific case $\mathbf{R}$ is the distance between the mass centers of the
tubes.  Using the fact that there are
two different time scales,
$\rho$ is expanded as
\begin{equation}
\rho(\mathbf{z},t)=\rho_0(\mathbf{z},\mathbf{R})+\epsilon
\rho_1(\mathbf{z},\mathbf{R},t)+\cdots
\label{expansion}
\end{equation}
where the first order term $\rho_0$ depends only on the slow time which is
regarded as the parameter $\mathbf{R}$.
The evolution of $\rho$ is given by the
Liouville equation. Injecting the expansion (\ref{expansion}) in the
Liouville equation gives a hierarchy of
equations which can be solved using a microcanonical distribution for
$\rho_0$. Having calculated $\rho$ to the first order
in $\epsilon$ one can find the following reaction force,
\begin{equation}
\mathbf{F}=-\mathbf{\nabla} E-\mathbf{\Gamma} \cdot\dot{\mathbf{R}}
\label{force}
\end{equation}
where the first order term is the Born-Oppenheimer force with the energy of the 
fast system $E$ and the second term the friction force where the velocity $\dot{\mathbf{R}}$ of order 
$\epsilon$ appears. The dynamic friction coefficients $\mathbf{\Gamma}$ are given by
\begin{equation}
\Gamma_{ij}=\frac{1}{\partial_E \Omega} \partial_E
\left( \partial_E \Omega \int_{0}^{\infty} d\tau \; C_{ij}(\tau) \right)
\label{gamma}
\end{equation}
where $\partial_E$ is the partial derivative with respect to the energy of the degrees of freedom 
of vibration $E$ and $\Omega(E,\mathbf{R})$ is defined by Eq. (\ref{invariant}). 
The autocorrelation function is given by the following expression
\begin{equation}
C_{ij}(\tau)=\mean{\partial_i \tilde{H}_{\tau} \partial_j
\tilde{H}}_{E,\mathbf{R}}
\label{correlation}
\end{equation}
where $\tilde{H}=H(z,\mathbf{R})-E(\mathbf{R})$ represents the fluctuations
of the Hamiltonian. We can simplify the
expression of the friction in Eq. (\ref{gamma}) by using the Boltzmann
equation $S=k_{\rm B} \ln \partial_E \Omega$ and the
thermodynamic relation $\partial E/\partial S=T$. Moreover, the adiabatic
parameter $\mathbf{R}$ is one-dimensional in our
system since the relative motion of the tubes is mainly in the longitudinal direction. We hence get
\begin{equation}
\Gamma=\left(\beta+\partial_{E}\right) \int_{0}^{\infty} d\tau \; C(\tau)
\approx \beta \int_{0}^{\infty} d\tau \; C(\tau)
\label{gammasimple}
\end{equation}
with $\beta=1/k_{\rm B}T$. In Eq. (\ref{gammasimple}), we have used the result that the second term with 
$\partial_E$ is a negligible correction of the order of the inverse of the system size. 
The autocorrelation function is given by
\begin{equation}
C(\tau)=\mean{F_{\rm LJ}(\tau)F_{\rm LJ}(0)}_{E,R}-\mean{F_{\rm LJ}}_{E,R}^2
\label{correlationsimple}
\end{equation}
where $F_{\rm LJ}$ is the total Lennard-Jones force in the transverse direction,
\begin{equation}
F_{\rm LJ}=\frac{1}{2}\sum_{ij} \frac{\partial V_{\rm LJ}}{\partial R}
\left(\mathbf{r}_i^{(1)}-\mathbf{r}_j^{(2)}\right)
\label{forcesimple}
\end{equation}
We observe that Eq. (\ref{gammasimple}) satisfies the
fluctuation-dissipation theorem relating the fluctuating
force acting on a system in a bath to the friction kernel in the
generalized Langevin equation:
\begin{equation}
k_{\rm B} T \; \zeta(t)=\mean{f(t)f(0)}
\label{frictionkernel}
\end{equation}
with $f=F_{\rm LJ}-\mean{F_{\rm LJ}}$.
Eq. (\ref{gammasimple}) is hence obtained in the Markovian limit where the
friction kernel (\ref{frictionkernel})
decays on a short time scale such that $\zeta(t)=\zeta_0\delta(t)$.

We are now going to show the results of the molecular-dynamics simulation.
We focus our numerical studies on a specific nanotube namely (5,0)@(15,0)
which is commensurate in the classification
of Ref. \cite{Crespi}.
The inner tube has $N_1=60$ atoms and a length of $l_1=10.9 \text{\AA}$.
The outer tube
has $N_2=240$ atoms and a length of $l_2=15.3 \text{\AA}$. The relative
mass is hence $\mu=576$ amu.
The molecular-dynamics simulation is straightforward. We first extract the
internal nanotube such as the initial distance
between both mass centers is $R_0=6.0 \text{\AA}$. The tubes centers of masses
have vanishing initial velocity and angular momentum.  We then start the
simulation and solve the Newtonian dynamics with
a fourth-order Runge-Kutta integration scheme in the microcanonical
ensemble. The initial temperature of the system is
$300K$, which is introduced by a Maxwell-Boltzmann distribution of the
initial velocities and a short early
time relaxation. We use a time step of $0.1$fs.

First, we can predict the energy to be dissipated by the sliding motion and
thus the increase of the temperature of the system.  Calculating the
difference of energies between the minimum of the
total Lennard-Jones potential $V_{\rm LJ}(R)$ and the initial value $V_{\rm
LJ}(R_0)$  gives us $\Delta E_s=0.623eV$. Using
$\Delta E= 3 N k_B T$ where $N=N_1+N_2$ we hence should have an increase in
temperature  of $\Delta T=8.04K$.

\begin{figure}
\begin{center}
\includegraphics*[scale=0.25,angle=0]{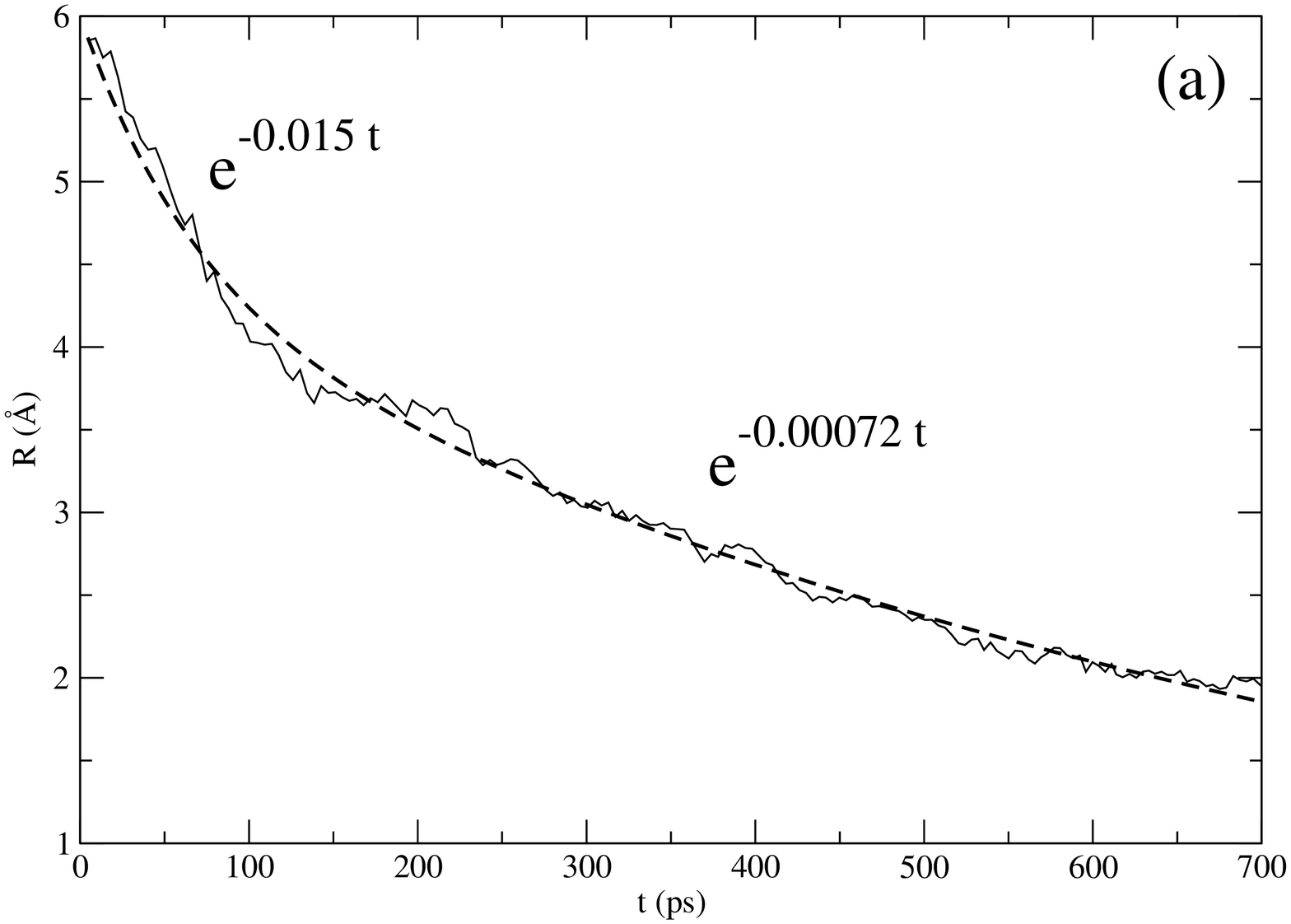}
\includegraphics*[scale=0.25,angle=0]{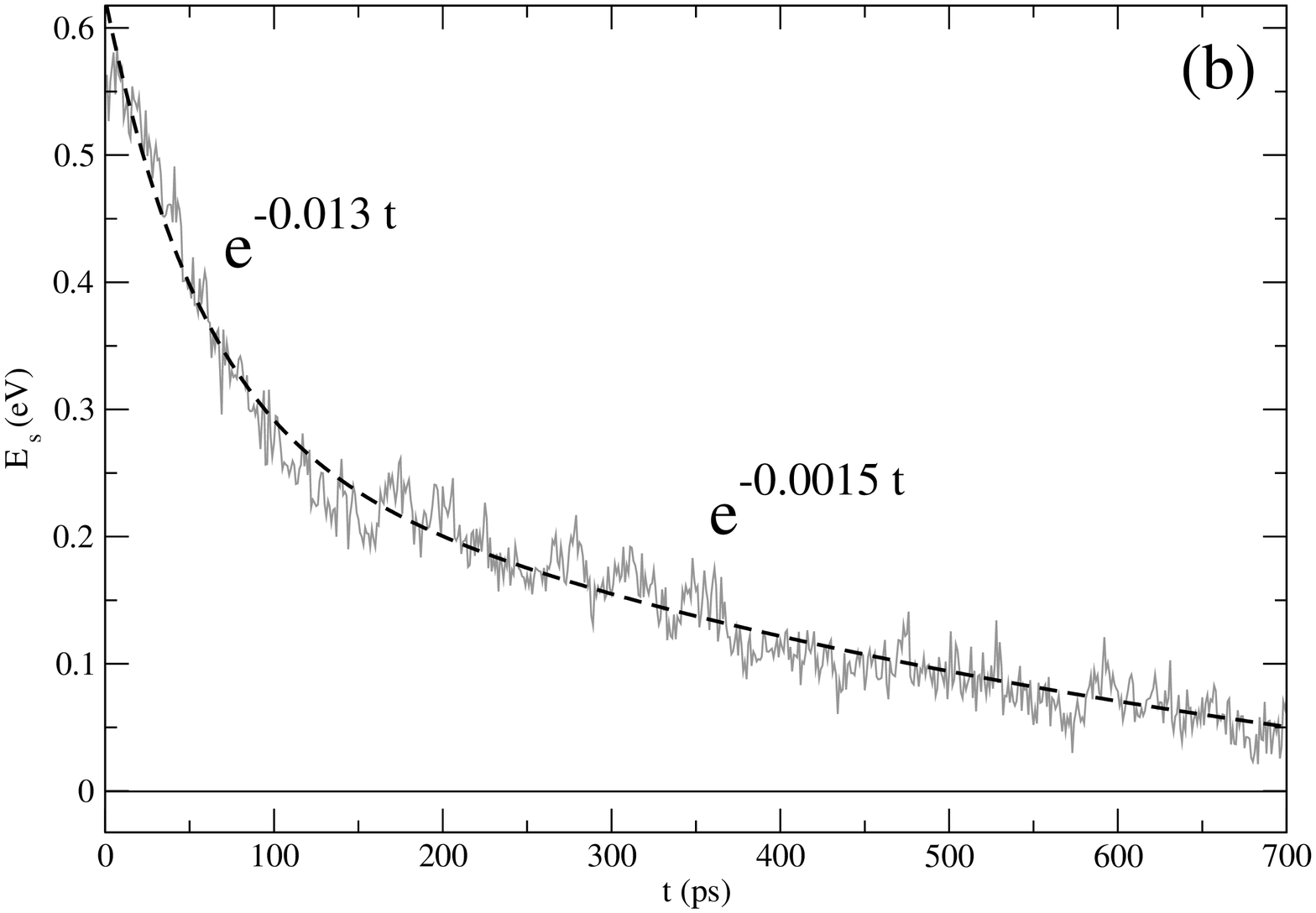}
\caption{(a) Evolution of the distance between mass centers represented by the
successive maxima at each oscillation. (b) Evolution of the energy of the relative 
motion also represented by the successive maxima at each oscillation. In this small 
system, the oscillations have a period of about 10ps. The crossover between the exponentials 
happens around 120fs when the bouncing of the edges ceases. The inner tube is entirely inside the 
outer one after about 400fs.}
\label{REevol}
\vspace{-0.7cm}
\end{center}
\end{figure}

We observe in Fig. \ref{REevol} a double exponential decay in the motion.
The origin of the double exponential decay holds in
the fact that, in the initial configuration, a part of the inner tube is
outside the outer tube. In this case, the
friction is higher because the inner tube interacts with the edge of the
outer tube and bounces during the reentry. It is
only after a  while that the inner tube oscillates inside the outer tube.
We see that the oscillations are damped as
expected.  We can use a simple model to describe the dynamics when the
inner tube remains inside, in which case the
potential energy of the relative motion between both tubes can be
approximated by a harmonic potential. We hence have a
Newton equation of the form,
\begin{equation}
\mu \ddot{R}=-\Gamma \dot{R}-kR
\label{newton}
\end{equation}
which can be solved to show that the successive maxima in position $R$ and
energy $E=\frac{1}{2}\mu \dot R^2 +\frac{1}{2} k R^2$ decay as
\begin{equation}
\vert R\vert_{\rm max}\propto e^{-\Gamma t/2\mu} \ \ \ \text{and}\ \ \
E_{\rm max}\propto e^{-\Gamma t/\mu}
\end{equation}
We hence expect to have a ratio of 2 between the damping rates of the
maxima in energy and position. By
performing fits of the curves in Fig. \ref{REevol}, we get
respectively $0.00072$/ps and $0.0015$/ps,
in good agreement with the expectation.
Multiplying by the relative mass $\mu$ we finally get the friction
coefficient $\Gamma=0.84 \pm 0.02$ amu/ps.

The next step is to calculate the friction coefficient with the
autocorrelation-function method. According to
Eqs. (\ref{gammasimple})-(\ref{forcesimple}), the friction coefficient is
given by the time integral of the autocorrelation
function of the force acting on the slow system with the distance between
the centers of mass fixed as a parameter. The
calculation of the autocorrelation function is also performed by
molecular-dynamics simulation except that the distance
between the centers of mass is kept constant by compensating at each time
step the motion of each atom by the motion of
the mass center of their respective nanotube.  We suppose that our system
is ergodic and mixing (meaning that its autocorrelation functions vanish asymptotically at long times). 
Since we have nearly one
thousand degrees of freedom this should be a good approximation. We can
thus use the ergodic theorem and replace the
ensemble average in Eq. (\ref{correlationsimple}) by a time average.

\begin{figure}
\begin{center}
\includegraphics*[scale=0.25,angle=0]{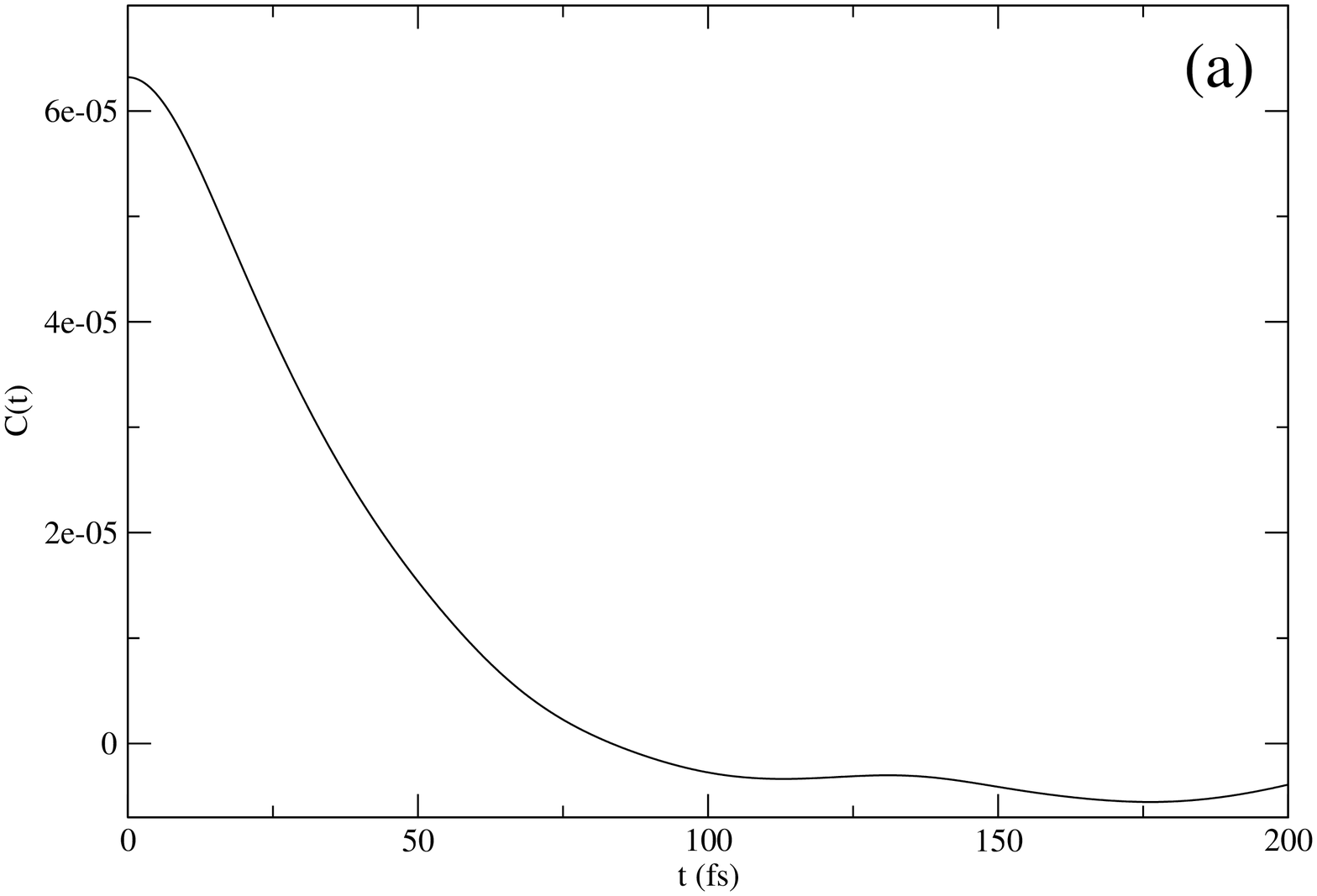}
\includegraphics*[scale=0.25,angle=0]{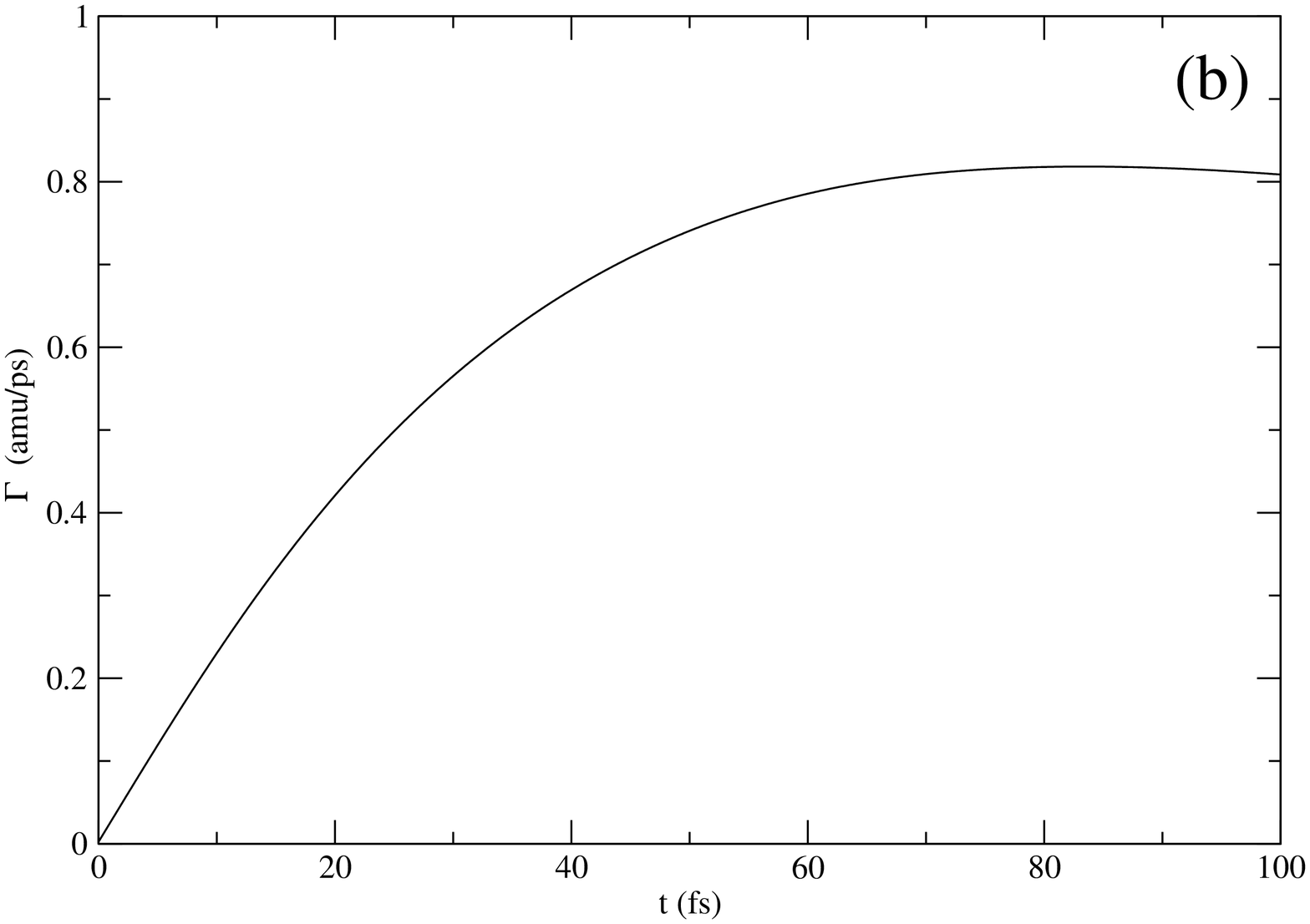}
\caption{(a) Autocorrelation function of the intertube force.
(b) Time integral of the autocorrelation function (a) numerically
converging to the friction coefficient. The small fluctuations seen beyond about 80fs in the 
autocorrelation-function (a) and its integral (b) give a negligible contribution and are reduced 
if the statistics used in the averaging are increased.}
\label{correl}
\vspace{-0.7cm}
\end{center}
\end{figure}

Figure \ref{correl}  depicts the results of
the numerical calculation of the autocorrelation-function 
and its time integral giving asymptotically the friction
coefficient. With this autocorrelation-function method,
we find the value $\Gamma=0.82$ amu/ps for the friction coefficient, in
excellent agreement with the value obtained
by direct molecular-dynamics simulation.  The great advantage of the
autocorrelation function method is that
the calculation is now about 50 times faster. Furthermore the
autocorrelation function method is much more
precise since it does not involve fitting to get the friction coefficient.

\begin{figure}[h!]
\begin{center}
\includegraphics*[scale=0.25,angle=0]{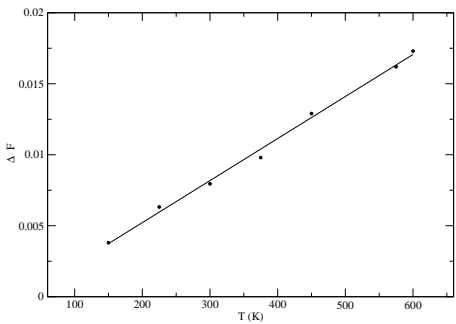}
\caption{Standard deviation of the force versus temperature.}
\label{deltaFT}
\vspace{-0.7cm}
\end{center}
\end{figure}

Furthermore, the autocorrelation-function method allows us to obtain a
theoretical formula for the friction
coefficient.  We first notice that the autocorrelation function decays on
the time scale of the vibrations
of the atoms in the carbon nanotubes.  This time scale is of the order of
the inverse of the Debye frequency
$\omega_{\rm D}$ so that the correlation time should be given approximately
by $t_{\rm C} \approx 2\pi /\omega_{\rm D}$.
Since the Debye temperature of the nanotube is $\Theta_{\rm D} \approx
1000$K \cite{Debye}, we hence expect the
autocorrelation function to decay to zero over a correlation time
approximately equal to 50fs, which is indeed confirmed by
Fig. \ref{correl}. Moreover, the autocorrelation function at time zero is
equal to the variance $\Delta F^2$ of
the force.  We can therefore approximate the time integral of the
autocorrelation function and obtained the following
approximate formula for the friction coefficient:
\begin{equation}
\Gamma \approx \frac{1}{k_{\rm B} T} \frac{1}{2} \frac{2\pi}{\omega_{\rm
D}} \Delta F^2
\end{equation}
Thanks to this formula, the evaluation of the friction coefficient reduces
to the evaluation of the variance of the
fluctuating force. On this ground, we now investigate the dependence of the
friction coefficient on temperature. We have
depicted in Fig. \ref{deltaFT} the result of the numerical calculation of
the standard deviation of the force.
We see that the standard deviation of the force increases linearly with the
temperature, hence the friction coefficient
also increases linearly with temperature: $\Gamma \sim T$. The fact that
friction increases with temperature in our system
can be understand by the fact that the dissipation of energy of the slow
motion is due to the coupling with the
vibrating degrees of freedom.  Increasing the temperature results into
larger vibrational motion and
thus into higher friction. Furthermore, the calculation shows that edge effects give 
to the small system we here considered, a larger friction than to a system of double length. 
Then the friction increases in systems from double to longer lengths, the coefficient remaining 
of the same order of magnitude as in the small system here studied.

In this work, we have developed theoretical and numerical methods to
calculate the dynamic friction coefficient
in the one-dimensional sliding motion of two concentric carbon nanotubes. We
focused our numerical studies on a commensurate
double-walled nanotube. We obtained by a molecular-dynamics simulation the
energy damping and thus the increase
in temperature of the system. The simulation shows the importance of edge
effects at early times: a significantly
larger damping is due to bouncing effects when the inner nanotube reenters.
This effect disappears at long times
when the inner tube moves inside the outer tube.  The long
time behavior is in good agreement with the one-dimensional model we
proposed. We have then calculated the friction with
the autocorrelation-function method. We observed that the autocorrelation
function decreases to zero with a time of the
order of 50fs which is a value in agreement with the experimental value of
the Debye temperature. The integral of the
autocorrelation function gives the same result as the direct
molecular-dynamics simulation with much less calculation time,
actually a factor 50, and with better precision since we avoid the use of
fitting methods. We can thus conclude that the
autocorrelation-function method is very efficient for the calculation of
friction coefficients.  Moreover, we have obtained
a formula which relates the dynamic friction coefficient to the Debye
frequency and the standard deviation
of the fluctuations of the forces between the nanotubes and determined that
the dynamical friction coefficient
increases with the temperature.

The Authors thank Pablo Jensen for helpful discussions. This research is
financially supported by the National Fund for
Scientific Research (FNRS Belgium).


\begin{thebibliography}{1}
\bibitem{Zettl}
J. Cumings and A. Zettl, Science \textbf{289}, 602 (2000)
\bibitem{Brenner}
D. W. Brenner, Phys. Rev. B \textbf{42}, 9458 (1990)
\bibitem{Zheng}
Q. Zheng and Q. Jiang, Phys. Rev. Lett. \textbf{88}, 045503 (2002)
\bibitem{Lu}
J. P. Lu and W. Yang, Phys. Rev. B \textbf{49}, 11421 (1994)
\bibitem{Lu2}
J. P. Lu, Phys. Rev. Lett. \textbf{79}, 1297 (1997)
\bibitem{Legoas}
S. B. Legoas, V. R. Coluci, S. F. Braga, P. Z. Coura, S. O. Dantas, and D. S. Galvao, Phys. Rev. Lett. \textbf{90}, 055504 (2003) 
\bibitem{Crespi}
A. N.  Kolmogorov and V. H. Crespi, Phys. Rev. Lett. \textbf{85}, 4727 (2000)
\bibitem{Jarzynski}
C. Jarzynski, Phys. Rev. Lett.  \textbf{71}, 839 (1993)
\bibitem{Berry}
M. V. Berry and J. M. Robbins, Proc. Roy. Soc. Lond. A. \textbf{442}, 659 (1993)
\bibitem{Ott}
E. Ott, Phys. Rev. Lett. \textbf{42}, 1628 (1979)
\bibitem{Goldstein}
H. Goldstein, C. Poole and J. Safko, \textit{Classical Mechanics}, third edition (Addison Wesley, 
Reading MA, 2002)
\bibitem{Debye}
J. Hone, B. Batlogg, Z. Benes, A.T. Johnson and J.E. Fisher, Science \textbf{289}, 1730 (2000)
\end{thebibliography}
\end{document}